\documentclass[showpacs,prl,floatfix,amsmath,amsfonts,superscriptaddress,twocolumn]{revtex4}
\usepackage{graphicx,amsmath,amsfonts,amssymb}
\usepackage[usenames]{color}
\usepackage{hyperref}
\newcommand{\pd}[2]{\frac{\partial #1}{\partial #2}}

\newcommand{\real}{\mathbb{R}}

\newcommand{\vcr}{v_\textrm{cr}}

\newcommand{\km}{k_\mathrm{max}}
\newcommand{\gm}{g_\mathrm{max}}
\newcommand{\vf}{v_\mathrm{fit}}

\newcommand{\p}{\boldsymbol{\psi}}

\begin{document}
\title{Dark-dark solitons and modulational instability in miscible,
       two-component Bose-Einstein condensates}
\author{M. A. \surname{Hoefer}}
\email{mahoefer@ncsu.edu} \affiliation{North Carolina State University, Department of Mathematics, Raleigh, NC 27695, USA}
\author{J.J. \surname{Chang}}
\author{C. \surname{Hamner}}
\author{P. \surname{Engels}}
\affiliation{Washington State University, Department of Physics and Astronomy, Pullman,
Washington 99164, USA}
\date{\today}

\begin{abstract}
  We investigate the dynamics of two miscible superfluids experiencing
  fast counterflow in a narrow channel. The superfluids are
  formed by two distinguishable components of a trapped dilute-gas
  Bose-Einstein condensate (BEC). The onset of counterflow-induced
  modulational instability throughout the cloud is observed and shown
  to lead to the proliferation of dark-dark vector solitons. These
  solitons, which we observe for the first time in a BEC, do not exist
  in single-component systems, exhibit intriguing beating dynamics and
  can experience a transverse instability leading to vortex line
  structures. Experimental results and multi-dimensional numerical
  simulations are presented.
\end{abstract}
\maketitle


Superfluids are a robust model system for the investigation of
nonlinear fluid flow.  Governed by an underlying macroscopic
wavefunction, superfluids can display a large variety of nonlinear
wave phenomena in the context of matterwaves.  In Bose-Einstein
condensates (BECs), nonlinear structures including solitons, vortices
and vortex rings have been the focus of intense research efforts
\cite{Dalfovo1999,Kevrekidis2009}. In this work, we investigate the
regime of fast counterflow between two distinguishable superfluids
in a narrow channel and observe dynamics leading to novel
structures. Modulational instability (MI), in which small
perturbations to a carrier wave, reinforced by nonlinearity,
experience rapid growth \cite{zakharov_modulation_2009}, plays a key
role in the dynamics. In many nonlinear systems, MI leads to the
breakup of periodic wavetrains, as in sufficiently deep water
\cite{benjamin_instability_1967}, as well as the formation of
localized structures in optics \cite{tai_observation_1986} and BECs
\cite{kevrekidis_2004}. In our case, MI-induced regular density
modulations, formed throughout the BEC, lead to the emergence of a
large number of \emph{beating dark-dark} solitons.  These
solitons--which exhibit periodic energy exchange between the two
condensate components \cite{park_systematic_2000}--are a
generalization of static dark-dark solitons
\cite{sheppard_polarized_1997}. They are distinctly different from all
previously observed solitons in BECs, including dark-bright solitons
which were generated in a two-component mixture by marginally
  critical counterflow-induced MI near a density edge
\cite{Hamner2010}.  We perform three-dimensional (3D) numerical
simulations to corroborate this interpretation and furthermore
identify a subsequent transverse instability resulting in
multi-dimensional structures such as vortex lines (see
\cite{Brand2002} for the scalar counterpart).

We study superfluid counterflow with an experimental system consisting
of BECs with typically $8 \times 10^5$ atoms of $^{87}$Rb. The BECs
are confined in a cigar shaped, far-detuned optical dipole trap with
measured trap frequencies of $2\pi\times$\{1.5, 140, 178\}~Hz with a
horizontal weakly confined axis.  By starting with all atoms in the
$|F, m_F \rangle =|1, -1\rangle$ hyperfine state and transferring
about 50\% of the atoms to the $|2, -2\rangle$ state via a 1~ms long
microwave sweep, a perfectly overlapped two-component mixture is
created. The predicted scattering lengths for these states
\cite{scattering_lengths} imply that this mixture is miscible
\cite{miscibility}, which is also supported by our experimental
observation of 
no phase separation for an unperturbed mixture of these states. To
induce relative motion between the components, an external magnetic
gradient is applied along the elongated (axial) direction. The
gradient pulls atoms in the $|2, -2\rangle$ state to the left and
those in the $|1, -1\rangle$ state to the right.  The atoms are imaged
using a free expansion imaging procedure. Each experimental image
shows an upper cloud consisting of the $|2, -2\rangle$ atoms after
7~ms of free expansion and a lower cloud consisting of the $|1,
-1\rangle$ atoms after 8~ms of free expansion. Both clouds are imaged
during the same experimental run.

\begin{figure}
  \centering
  \includegraphics[width=0.9\columnwidth]{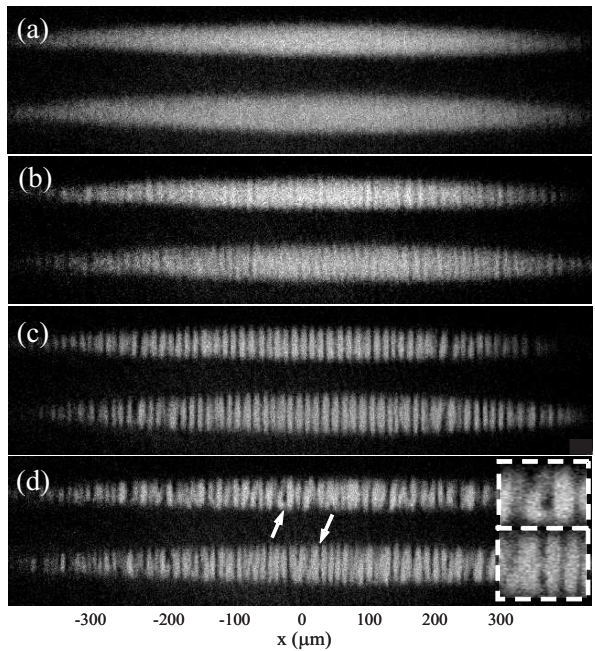}
  \caption{Counterflow induced MI in the presence of a strong magnetic
    gradient of 10.4~mG/cm.  Evolution times (a)~10~ms, (b)~70~ms,
    (c)~95~ms. (d) After MI onset, the axial gradient is turned off,
    followed by a trapped evolution time of 20~ms.}
  \label{fig:mi_onset}
\end{figure}

Experimental data showcasing the formation of a very dense
counterflow-induced MI pattern are presented in
Fig.~\ref{fig:mi_onset}. In the presence of a 10.4~mG/cm axial
gradient, gradual pattern formation starts after 70~ms of smooth
evolution (Fig.~\ref{fig:mi_onset}(a, b)). We first observe pattern
formation in non-central regions where the two condensates have
differing densities (Fig.~\ref{fig:mi_onset}(b)).  This is due to the
dependence of the critical velocities for counterflow-induced MI on
the two component density ratio, being largest when the densities are
equal \cite{law_critical_2001,Hamner2010}. 
After 
about 25~ms, 
a very dense and regular MI pattern 
fully develops, filling the entire BEC
(Fig.~\ref{fig:mi_onset}(c)). The modulations in the two components
are offset in the axial direction in a staggered way such that
  one component fills the depressions in the other.  Under the
continued influence of the axial gradient, the regular pattern of
Fig.~\ref{fig:mi_onset}(c) quickly becomes uneven and irregular.
Alternatively, if the gradient is switched off after the MI pattern
has fully developed, we frequently observe the formation of black dots
such as those marked by the arrows in Fig.~\ref{fig:mi_onset}(d) which
might indicate the generation of vorticity.  We note recent
theoretical work suggesting that counterflow-induced MI 
may be used to generate quantum turbulence \cite{tsubota2010}.

Imparting slow counterflow conditions, implying slow MI onset in
  the quasi-uniform background, we previously generated a dark-bright
soliton train emanating \emph{locally} from a density edge
\cite{Hamner2010}.  In contrast, the fast counterflow considered here
leads to rapid MI onset and pattern formation \emph{throughout}
the condensates.

\begin{figure}
  \centering
  \includegraphics[width=0.9\columnwidth]{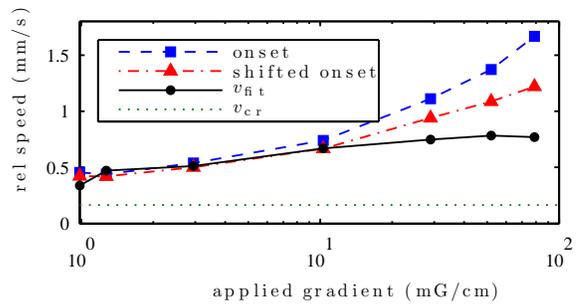}
  \caption{(Color online) Consistent predictions of counterflow speed
    based on wavelength and onset time measurements of MI as a
    function of applied gradient. For details see text.}
    \label{fig:relspeeds}
\end{figure}

MI theory agrees quantitatively with the experimentally observed
patterns as we now explain [Fig.~\ref{fig:relspeeds}].  For a uniform
counterflow, the onset of MI corresponds to a complex sound speed (see
\cite{EPAPS}) and exhibits a preferred wavenumber, $\km$,
corresponding to the maximum growth rate $\gm$, both depending on the
counterflow speed.  
Unfortunately, our imaging procedure does not allow us to determine
the counterflow speeds experimentally. However, we can take two
independent theoretical approaches, described below, to extract the
onset velocities from our experimental data. The fact that these two
independent approaches lead to consistent results gives quantitative
credence to the theory. First, we use the analytical theory in
\cite{law_critical_2001,Hamner2010} to calculate the counterflow speed
$\vf$ whose corresponding $\km$ equals the experimentally observed
pattern periodicity at the trap center where the densities are assumed
equal (solid, black curve in Fig.~\ref{fig:relspeeds}).  In a second,
independent approach, we assume spatially uniform counterflow whereby
the applied gradient leads to unimpeded acceleration of each component
(calculated from the atomic magnetic moment and the magnitude of the
applied gradient).  Using this simple model, experimentally determined
onset times for MI are converted to relative speeds at the onset of
the MI pattern (dashed blue curve in Fig.~\ref{fig:relspeeds}).  The
dash-dotted red curve in Fig.~\ref{fig:relspeeds} uses the same,
uniform counterflow model but shifts the measured MI onset time by
$-1/\gm$.  Subtracting this time accounts for the development of the
instability and leads to a better approximation of the true relative
speed that sets the pattern periodicity.  The resulting curve
interpolates the two models.  The lowest, dotted curve is the
predicted critical speed in the condensate center ($\vcr = 0.16$ mm/s)
demonstrating fast counterflow.  Despite the approximations made, the
curves exhibit agreement for small to moderate gradients, suggesting
that the observed dynamics are theoretically described by
counterflow-induced MI.  Discrepancies at large gradients are likely
due to the large accelerations involved and spatial nonuniformity.

\begin{figure}
  \centering
  \includegraphics[width=0.9\columnwidth]{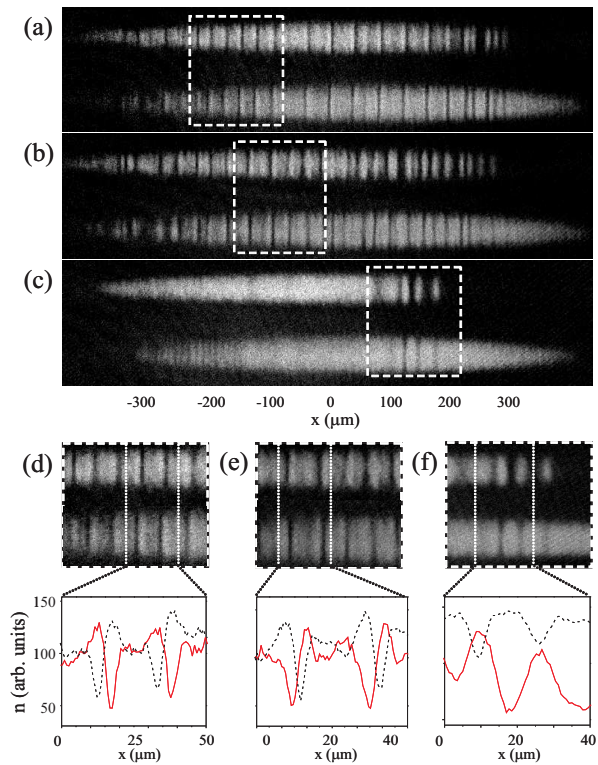}
  \caption{(Color online) (a, b) Dark-dark solitons as a result of MI
    after applying a gradient of 1.4~mG/cm for 350~ms and 380~ms,
    respectively. (c) Formation of dark-bright solitons when a small
    magnetic gradient of 0.2~mG/cm is applied. The gradient is left on
    for 1000~ms before the start of the expansion sequence.  (d-f):
    Zoomed-in view of boxed region in (a-c), respectively.  Red solid
    lines are integrated cross sections of $|2, -2\rangle$ state,
    black dashed lines of $|1, -1\rangle$ state.  }
  \label{fig:smallgradient}
\end{figure}

We now investigate the dynamics of the MI onset by using a smaller
gradient of 1.4~mG/cm so that $\km$ is reduced relative to
Fig.~\ref{fig:mi_onset}, enabling better experimental observation of
individual features (Fig.~\ref{fig:smallgradient}). After 
smooth counterflow, MI sets in across the BEC 
leading to a 
regular array of dark-dark solitons (Fig.~\ref{fig:smallgradient}(a,
b, d, e)). In accordance with theory and our numerics (see below), the
dark-dark solitons exhibit a dynamic beating as seen by comparing the
integrated cross sections of Fig.~\ref{fig:smallgradient}(d,e), noting
the order of the notch and bump feature in each component. While our
destructive imaging technique does not allow us to determine the exact
beat frequency, our 3D numerics indicate a timescale of fifteen
milliseconds per period \cite{EPAPS}. The dark-dark solitons we
observe here are new and distinct from the dark-bright solitons that
have been observed previously in BECs
\cite{Anderson2001,Becker2008,Hamner2010}, 
being distinguished by their far field conditions and dynamics.  To
facilitate a comparison, an example dark-bright soliton train, seeded
at the condensate interfaces and generated by slow, marginally
critical counterflow \cite{Hamner2010}, is shown in
Fig.~\ref{fig:smallgradient}(c, f).  A dark-bright soliton consists of
a dark notch in one component, filled by a localized density bump of
the second component. In contrast, the beating dark-dark soliton
asymptotes to nonzero densities in both components and dynamically
changes its shape, with each component possessing a density bump
adjacent to a notch which alternate their relative positions in time
[see also Fig.~\ref{fig:beatingmodel} below].

\begin{figure}
  \centering
  \includegraphics[width=0.9\columnwidth]{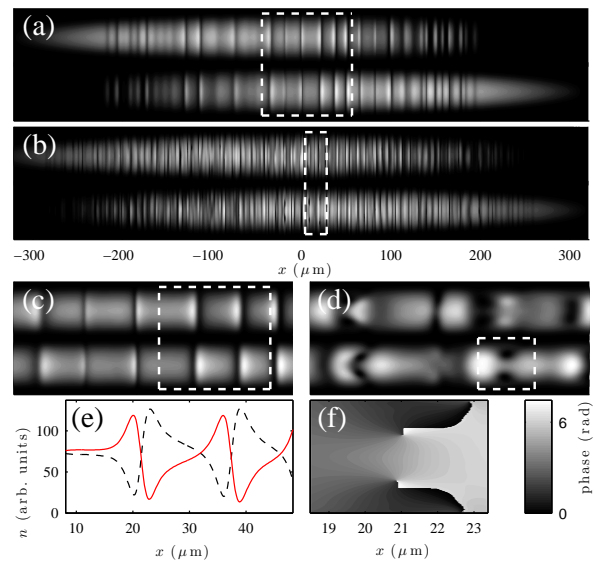}
  \caption{
    (Color online) Integrated densities from 3D numerical simulations.  (a,c,e)
    correspond to Fig.~\ref{fig:smallgradient}(a,b) at $t = 421$ ms
    with zoom in (c) and line plot (e) of dark-dark solitons.  (b,d,f)
    correspond to Fig.~\ref{fig:mi_onset}(a-c) at $t = 133$~ms with
    zoom in (d) and a phase plot along the vertical $z =
    0$ plane (f) showing two vortex
    lines with oppositely oriented $2\pi$ phase winding.  The vertical
    axes of (a-d) span 16.7~microns incorporating a
    vertical offset of 8 microns between the clouds.
  }
  \label{fig:mi_1d_sim}
\end{figure}

The dynamics are well reproduced by 3D numerical simulations
\cite{EPAPS} of the vector, mean-field Gross-Pitaevskii equation with
initial conditions and parameters corresponding to the experiments in
Figs.~\ref{fig:mi_onset}(a-c) and \ref{fig:smallgradient}(a,b). As
with experiment, a smooth, accelerating counterflow develops due to
the axial field gradient.  Dark-bright solitons form at the edges of
the condensates until the rapid growth of large scale modulations is
observed (Fig.~\ref{fig:mi_1d_sim}(a,b)).  For moderate gradients in
Figs.~\ref{fig:mi_1d_sim}(a,c,e), 
these modulations rapidly develop into a number of localized,
essentially one-dimensional (1D) beating dark-dark solitons with
initial approximate spacing $2\pi/\km$. 
Continued evolution results in 
interactions and eventual solitary wave transverse
breakup 
at about $t = 600$ ms.

For the strong 
gradient case,
our numerics show the development of axial modulations by about $t =
125$ ms with an initial 1D structure.  In contrast to the moderate
gradient regime, these structures \emph{rapidly} undergo decay due to
transverse modulations which leads to the formation of columnar
2D vortex lines, Fig.~\ref{fig:mi_1d_sim}(b,d), exhibiting a $2\pi$
phase winding around their core, Fig.~\ref{fig:mi_1d_sim}(f), and a
uniform structure along the direction of view.  The numerics
also show vortex lines oriented along the orthogonal,
horizontal 
radial axis.  In analogy to the scalar case \cite{Brand2002}, we
interpret this behavior as a transverse instability that depends on
the relative speeds of the two components, their densities, and the
transverse confinement strength.  

\begin{figure}
  \centering
  \includegraphics[width=\columnwidth]{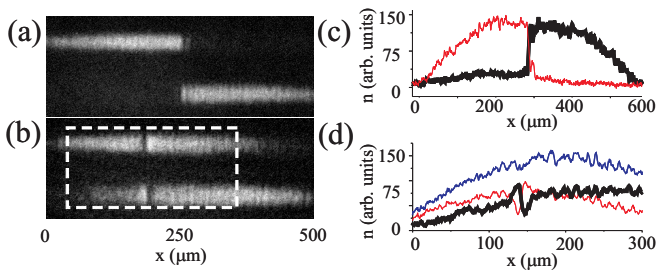}
  \caption{(Color online) Dark-dark solitons as a result of two
    component mixing. (a) Phase separated mixture in presence of axial
    gradient. (b) Dark-dark soliton formed 1~sec after sudden turn-off
    of gradient.  (c, d) Integrated cross sections with thin red
    (thick black) curve showing the $|2,-2\rangle$ ($|1,-1\rangle$)
    component. Blue (upper) curve in (d) shows total density.}
  \label{fig:mixing}
\end{figure}

\begin{figure}
  \centering
  \includegraphics[width=0.9\columnwidth]{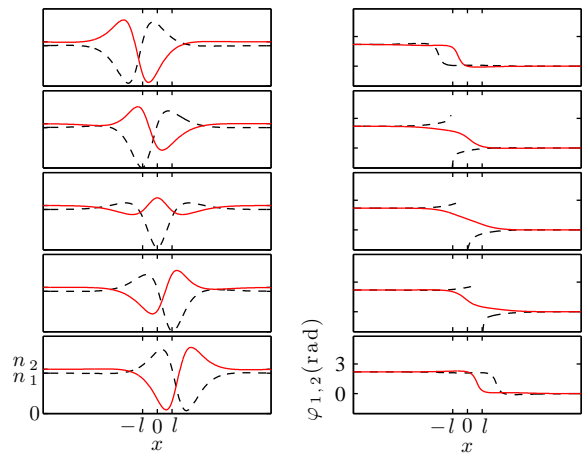}
  \caption{(Color online) Density and phase evolution of a beating
    dark-dark soliton assuming equal scattering lengths.}
  \label{fig:beatingmodel}
\end{figure}

Dark-dark solitons can also be observed in other settings, e.g.
during the mixing of two initially phase separated components. An
experimental result is showcased in Fig.~\ref{fig:mixing}. We start
from the phase separated situation in Fig.~\ref{fig:mixing}(a,c) which
forms after initially overlapped components experience an axial
gradient for 10~sec. When the axial gradient is suddenly switched off,
the two components interpenetrate, first forming a smooth and extended
overlapped region. After some evolution time, individual dark-dark
solitons appear (Fig.~\ref{fig:mixing}(b, d)) exhibiting approximately
constant total density (upper, blue curve).  This behavior is
reminiscent of dark soliton formation in colliding single-component
BECs \cite{Scott1998}.  Beating dark-dark solitons are also
theoretically predicted to develop when a repulsive beam is swept
through a two-component miscible BEC with an appropriate speed
\cite{Kevrekidis2007}.


The beating solitons can be understood through the following
simplified model: assuming that all scattering lengths are equal to
$a_{22}$, the 
mean field equation is the repulsive, vector Nonlinear Schr\"{o}dinger
(NLS) equation.  Its most general known soliton solution is the six
parameter dark-dark soliton \cite{park_systematic_2000} (e.g.~two
background densities $n_{1,2}$, two background flow speeds $c_{1,2}$,
soliton speed $v$, and beating frequency $\omega$) of which the
well-studied five parameter static dark-dark soliton
\cite{sheppard_polarized_1997} is a special case.  Even though
analytical expressions for these solitons were derived
\cite{park_systematic_2000}, their form is quite complicated and basic
properties such as the beating frequency as a function of soliton
parameters are unknown.

An 
example of a beating dark-dark soliton can be constructed by
leveraging the SU(2) invariance of the vector NLS equation
\cite{EPAPS}.  Applying a rotation matrix to the two components of a
four parameter dark-bright soliton \cite{sheppard_polarized_1997}, we
obtain a five parameter beating dark-dark soliton where the background
flow speeds are equal to $c$.  Its evolution over half a beating
period is shown in Fig.~\ref{fig:beatingmodel} (compare with
Figs.~\ref{fig:smallgradient}(d,e) and \ref{fig:mi_1d_sim}(e)).  The
beating angular frequency $\omega = \frac{m}{2 \hbar}(c-v)^2 \sec^2
(\phi/2)$ satisfies \cite{EPAPS}
\begin{equation}
  \label{eq:1}
  m(c-v)^2/(2\hbar) < \omega < \pi \hbar a_{22} (
  n_1 + n_2)/m . 
\end{equation}
The soliton half-width is 
$l = \hbar/\sqrt{2 m \omega \hbar - m^2(c-v)^2}$, where 
$\phi$ is the soliton phase jump and $m$ is the particle mass.  As
$\omega$ approaches the lower (upper) bound in (\ref{eq:1}), the
beating soliton degenerates to a plane wave (static dark-dark
soliton).  The beating soliton strongly resembles features observed in
experiment and numerical simulations.  The predicted minimum
oscillation period of 5 ms for our experimental parameters is
consistent with the numerically observed periods of about 15 ms.

In conclusion we have presented the first experimental observation of
a beating dark-dark soliton. 
These solitons naturally arise from a fast counterflow-induced
modulational instability and 
can emerge during the mixing of two superfluids.  Our work opens the
door to a range of new studies of vector soliton dynamics, with
consequences for a diversity of nonlinear, dispersive systems.

\acknowledgments{P.E. acknowledges support from NSF and ARO.  M.A.H.
  acknowledges support from NSF DMS1008973.}


\bibliographystyle{apsrev}

\newpage

\begin{widetext}
  
\section{Supporting material for ``Dark-dark solitons and modulational
  instability in miscible, two-component Bose-Einstein condensates''
  by M. A. Hoefer, J. J. Chang, C. Hamner, and P. Engels:
  numerical simulations}

\subsection{Numerical Computations}
\label{sec:numer-comp}

Our 3D numerics are performed on the vector Gross-Pitaevskii equations
\begin{equation}
  \label{eq:1}
  \begin{split}
    i \hbar \pd{\Psi_1}{t} &= -\frac{\hbar^2}{2m} \Delta \Psi_1 +
    V_1(\mathbf{x},t)
    \Psi_1 + (g_{11} |\Psi_1|^2 + g_{12} | \Psi_2|^2) \Psi_1, \\
    i \hbar \pd{\Psi_2}{t} &= -\frac{\hbar^2}{2m} \Delta \Psi_2 +
    V_2(\mathbf{x},t)
    \Psi_2 + (g_{21} |\Psi_1|^2 + g_{22} | \Psi_2|^2) \Psi_2 , \\
    \int_{\real^3} &| \Psi_j(\mathbf{x},t) |^2 d\mathbf{x} = N_j, \quad g_{jl} =
    \frac{4\pi \hbar^2 a_{jl}}{m}, \quad 1 \le j,l \le 2 .
  \end{split}
\end{equation}
where $g_{jl} = 4\pi \hbar^2 a_{jl}/m$, $a_{jl}$ are the $s$-wave
scattering lengths, $m$ is the particle mass, $N_j$ are the number of
condensed atoms in each component, and
\begin{equation}
  \label{eq:2}
  V_j(\mathbf{x},t) = \frac{1}{2} m (\omega_x^2 x^2 + \omega_y^2 y^2 +
  \omega_z^2 z^2) 
  + \mu_B g_j m_j B'(t) z, \quad j=1,2 ,
\end{equation}
with the time varying magnetic field gradient $B'(t)$, $\mu_B$ is the
Bohr magneton, the Land\'{e} g-factor is $g_j$ which has opposite
signs for each component $g_1 = -1/2$ and $g_2 = +1/2$, hence field
gradient induced counterflow.  The hyperfine quantum number is $m_j$.

We use the split-step pseudospectral Fourier method \cite{Bao2003}
(adapted to multiple components) with a uniform grid spacing of
approximately 0.19~$\mu$m in a rectangular 11.7~$\mu$m $\times$
11.7~$\mu$m $\times$ 772~$\mu$m box with periodic boundary conditions
and a time step of approximately 0.0089~ms. The initial condition is
computed by iteration of the stationary equations leading to the
ground state in the presence of the time-independent portion of the
trap.  The experimentally invoked free expansion directly before
imaging the condensate was not performed in the numerical
simulations. As is common with the numerical simulation of unstable
systems, we find that the onset time of modulational instability
depends on the amount of noise in the system.  Therefore, we construct
noise with zero mean, $10^{-4}$ variance, Gaussian distributed Fourier
components for the smallest 32 wavenumbers along each spatial
dimension.  $(1 + \textrm{noise})$ multiplies the initially smooth
density and is used as the initial condition.

The movies contained in the EPAPS material show results obtained from
our 3D numerics, simulating the experimental procedure of Figs.~1 and
3 of the main text. The files
\texttt{contour\_strong\_gradient\_large.mov} and
\texttt{contour\_strong\_gradient\_zoom\_large.mov} correspond to the
experiments in Fig.~1(a-c) and the numerical simulations in
Fig.~4(b,d,f) of the main text.  The file
\texttt{contour\_moderate\_gradient\_large.mov} corresponds to the
experiments in Fig.~3(a,b,d,e) and the numerical simulations in
Fig.~4(a,c,e) in the main text.  In all cases, smooth counterflow
leads to the spontaneous generation of a number of beating dark-dark
solitons and their decay due to transverse modulations.

\subsection{Beating Dark-Dark Soliton}
\label{sec:beating-dark-dark}

The two-component vector Nonlinear Schr\"{o}dinger equation modeling
the (1+1)d dynamics of untrapped, binary BECs with equal scattering
lengths is
\begin{equation}
  \label{eq:3}
  i \p_t = -\frac{1}{2} \p_{zz} + \| \p \|^2 \p , \quad \p =
  (\psi_1,\psi_2)^T .
\end{equation}
This equation can be obtained from eq.~(\ref{eq:1}) by a suitable
dimensional reduction (see e.g.~\cite{Kevrekidis2009}) and
nondimensionalizing lengths by the transverse harmonic oscillator
length $a_0 = \sqrt{\hbar/m \omega_x}$, time by the transverse trap
frequency $\omega_x$, and density by $2\pi a_s a_0^2$ where we have
assumed that all scattering lengths are equal to $a_s$. Four parameter
dark-bright soliton solutions are well-known
\cite{sheppard_polarized_1997}.  Because eq.~(\ref{eq:3}) exhibits
SU(2) invariance, we can ``rotate'' the most general dark-bright
soliton and obtain a five parameter beating dark-dark soliton which
was discussed in \cite{park_systematic_2000}
\begin{equation}
  \label{eq:4}
  \begin{split}
    \psi_1 = & \sqrt{\rho_1} \left \{ \cos \phi + i \sin \phi
    \tanh[a(z-vt)] \right \}
    \exp \left \{ i[cz - (c^2/2+\rho_1 + \rho_2) t] \right \} - \\
    & \sqrt{\rho_2 \sin^2 \phi - a^2 \frac{\rho_2}{\rho_1 + \rho_2}}
    \, \mathrm{sech}[a(z-vt)] 
    \exp \left \{i[vz + (a^2/2 - v^2/2 - \rho_1 - \rho_2 )t]  \right \}, \\
    \psi_2 = & \sqrt{ \rho_2} \left \{ \cos \phi + i \sin \phi
    \tanh[a(z-vt)] \right\}
    \exp \left \{ i[cz - (c^2/2+\rho_1 + \rho_2) t] \right \} + \\
    & \sqrt{\rho_1 \sin^2 \phi - a^2 \frac{\rho_1}{\rho_1 + \rho_2} }
    \, \mathrm{sech}[a(z-vt)] \exp \left 
      \{i[vz + (a^2/2 - v^2/2 - \rho_1 - \rho_2 )t] \right \} .
  \end{split}
\end{equation}
The soliton parameters are the far field densities $|\psi_j|^2 \to
\rho_j$, $|z| \to \infty$, the soliton inverse half-width $a$, the
phase jump $2\phi$ is the same for each component, the soliton speed
$v$, the background flow speed $c$ which is also the same for each
component, and the beating frequency $\omega$ due to the relative,
time dependent phases multiplying the $\tanh$ and $\mathrm{sech}$
terms.  There are five independent parameters with the other two, say
$a$ and $\phi$, related through
\begin{equation}
  \label{eq:5}
  \cos \phi = \frac{c-v}{\sqrt{2 \omega}}, \quad a = \sqrt{2 \omega -
    (c-v)^2} .
\end{equation}
Due to parameter restrictions on the original, un-rotated dark-bright
soliton, the dark-dark soliton exists only for beating frequencies in
the range
\begin{equation}
  \label{eq:6}
  \frac{(c-v)^2}{2} <
  \omega < \frac{\rho_1 + \rho_2}{2} .
\end{equation}
One can directly verify that the beating dark-dark soliton in
\eqref{eq:4} bifurcates from a plane wave solution at $\omega =
(c-v)^2/2$ and a static dark-dark soliton for $\omega = (\rho_1 +
\rho_2)/2$.

\subsection{Sound Speeds}
\label{sec:sound-speeds}

\begin{figure}[h]
  \centering
  \includegraphics[width=8.5cm]{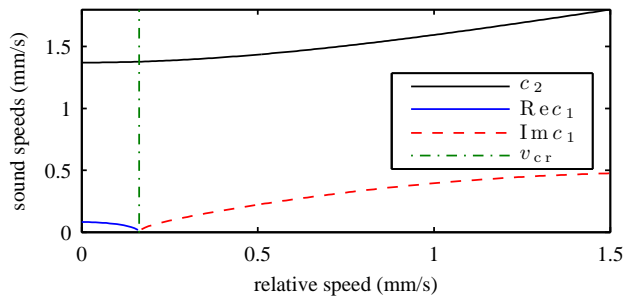}
  \caption{Sound speeds as functions of the relative speed.  The
    smallest speed becomes purely imaginary for relative speeds above
    critical $v_\text{cr}$, indicating the onset of modulational
    instability.}
  \label{fig:sound_speeds}
\end{figure}
For completeness, we include a plot (Fig.~\ref{fig:sound_speeds}) of
the axial sound speeds \cite{law_critical_2001}
\begin{equation}
  \label{eq:7}
  c = \pm \left [ \frac{2 \pi a_{22} | \Psi_2 |^2 \hbar^2}{m^2} +
    \frac{1}{4} v_\text{rel}^2 \pm   \left ( \frac{2 \pi a_{22} |
        \Psi_2 |^2 \hbar^2 v_\text{rel}^2}{m^2} + \frac{4 \pi^2
        a_{12}^2 a_{22} |
        \Psi_2 |^4 \hbar^4}{a_{11} m^4} \right )^{1/2} \right ]^{1/2},
\end{equation}
for the case when the densities and scattering lengths satisfy the
relation
\begin{equation}
  \label{eq:8}
  a_{11} | \Psi_1 |^2 = a_{22} | \Psi_2 |^2 ,
\end{equation}
which corresponds to the experiments and analysis in Fig.~2 of the
main manuscript.  The relative superfluid speed is
\begin{equation}
  \label{eq:9}
  v_\text{rel} = \frac{\hbar}{m} | (\arg \Psi_2)_z - (\arg \Psi_1)_z | .
\end{equation}

\end{widetext}

\end{document}